\documentclass[12pt]{revtex4}
 \usepackage{graphics, graphicx}
 \usepackage{color}
 \begin{document}

 
 \title{Working With Stephen}
 
 \author{James B.~Hartle}

\email{hartle@physics.ucsb.edu}

\affiliation{Santa Fe Institute, Santa Fe, NM 87501}
\affiliation{Department of Physics, University of California,
 Santa Barbara, CA 93106-9530}


\begin{abstract} 
The banquet for  the July 2017 conference in Cambridge, UK  celebrating Stephen Hawking's 75th birthday was held in Trinity College on July 3rd.   The organizers asked the author, among others,  to give a 10 minute after dinner talk on what it was like to work with Stephen. The following is an edited version of the author's speaking text. 
\end{abstract}

\maketitle

 \large
 
 I would like to thank the organizers for inviting me to this  celebration and for giving me an opportunity to speak at a wonderful occasion. For after dinner amusement they asked me to give a few personal recollections addressing the question `What is it like to work with Stephen'. It remains to be seen whether I should thank them for that.
 
 My association with Stephen began began 46 years  ago during a long visit to Fred Hoyle's Institute of Theoretical Astronomy (as it was known then). In residence were people like Brandon Carter, Martin Rees, Paul Davies, and Stephen Hawking --- colleagues with whom I have maintained lifelong personal and scientific contacts.  
 
 Stephen had guessed there must be solutions of the Einstein-Maxwell equations that represented many equally charged equally massive black holes held in equilibrium with the electrostatic repulsion  between them balancing their gravitational attraction. As it happened I knew where the relevant metric was to be found, and we were off and running.  
 
  \begin{figure}[t]
\includegraphics[width=6in]{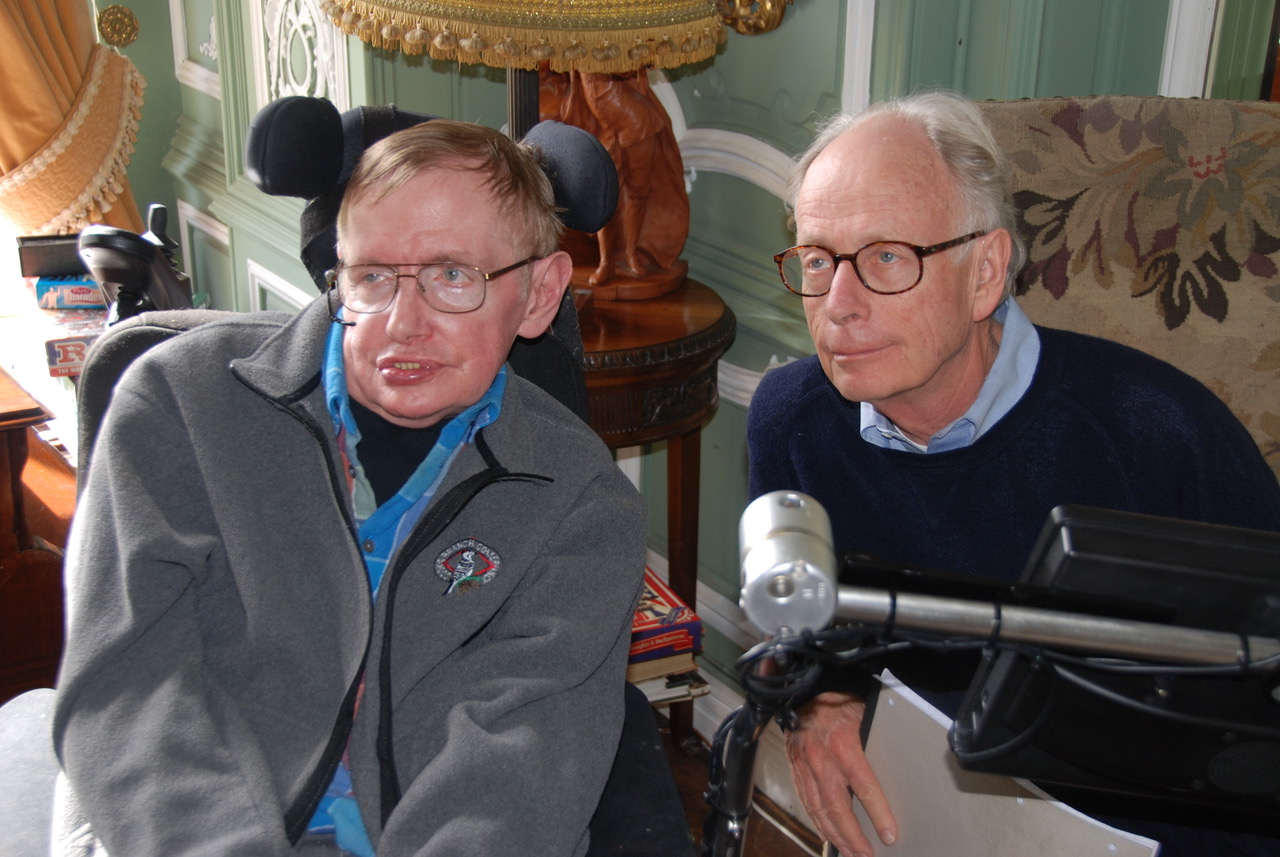}
\caption{The author and Stephen Hawking working .    (Photo courtesy Kathy Page)}
\label{swh-jbh}
\end{figure}

 Since that time I have always  felt that we were on the same wavelength --- not the same in ability or insight,  of course --- but rather similar in style and understanding of what is important. Ten more papers were to follow. 
 
 Its impossible to address the question of what is it like to work with Stephen without first answering the question of  what is it  that makes him so special as a scientist. 
 
 Its a common misconception that scientists working at the frontier are participants in an organized effort  seeking to answer some well defined fixed famous great questions. ``What is the unified theory of all the fundamental interactions?''  ``What was the origin of the universe?''  But this is not how it seems to those of us  in the  trenches.  

The frontier of theoretical physics is a noisy, chaotic place. At any one time, we have cherished {\it old}\  ideas, well confirmed in territory already mapped, that we are very reluctant to give up. We also have a whole variety of competing, different, unconfirmed {\it new} ideas vying for the best route forward into {new} territory that {\it require} us to give up some of the old ideas. It's certainly not organized! I think that Stephen likes situations like this and the surprises that upset the status quo.

Rather the questions are: ``What is the  right question?''  ``What do we keep and what do we give up?'' Stephen knows the answers. 
He  has a remarkable ability to see through all this clutter, to cut to the heart of  the matter, and to focus on the essentials. He also has the courage to discard the cherished old ideas that are an obstacle to progress. Later when looked at in the right way these seem inevitable. But that's the genius. 

Working with Stephen day to day is not so very different from other collaborations. Ideas are supplied and vetted by all. Contrary views are reconciled by argument. The important is distinguished from the unimportant by debate. The feasible is separated from the unfeasible by trial and error.  Stephen is inclusive in all this and I think cognizant of its value. Many of his statements start with `we', for instance: `We could say that...'  That inclusion extends to a larger loose group of scientists with similar viewpoints  working on similar problems --- Thomas Hertog for instance, Stephen's many students,  and many more. Working with Stephen broadens your horizons both scientifically and personally. 

Working with Stephen is fun. You get to draw on his vast and deep understanding of physics broadly and general relativity in particular. You learn. You get to appreciate his ability to explain with concision and clarity. You get to enjoy Stephen's wonderful sense of humor.  Right now I can see the expression on his face when he getting ready to tell me that something I've just said is inconsistent, or violates some basic principle of physics, or some other rubbish. But it's not just about physics.  The conversation is often diverted to talk about other issues in science and life on which Stephen often has strong, well thought out positions. 

The consequences of working with Stephen extend far beyond the particular paper or calculation.   I have often thought that the signature of a great problem in physics is that its solution generates more great problems. Certainly that is the case with the Hawking radiation. In my case the work with Stephen on the no-boundary wave function of the universe led to numerous specific calculations, many with Thomas Hertog and Stephen, of what it predicts for our observations of the universe on the largest scales of space and time.   But it also motivated a new vision formulated with Murray Gell-Mann  of how  how usual text-book quantum mechanics can be generalized to apply to cosmology --- decoherent histories quantum theory in particular.  

Working with Stephen brings out the best in you.   You are stretched.  You learn. You are confronted with new ideas, and new perspectives that are more powerful than the ones you had before.  Important new routes are opened up and important questions asked that you didn't think existed.  Stephen encourages thinking outside the box while at the same time  setting  high standards for that thinking.    

Late in life I find that I am proud of my record in theoretical physics. Not, of course,  because of any great discoveries or huge impacts.   But rather in the sense of `personal best'. I think that I did the best that I could have done with the talents and circumstances that I had.  But I don't think I could have done even that without working with  Stephen.

\vskip .3in 

A toast concluded the
 proceedings:  ``For seventy-five years of achievement both in science and life:   Stephen Hawking''. 
 
 \vskip .5in
 
 {PAPERS WITH STEPHEN HAWKING}
 
 \begin{itemize}
 
\item {}  Solutions of the Einstein-Maxwell Equations with Many Black
Holes, {\sl  Comm. Math Phys.}, 26, 87-101, 1972.

\item {}  Energy and Angular Momentum Flow into a Black Hole, {\sl Comm. Math. Phys}., 27, 283-290, 1972.

\item {}  Path Integral Derivation of Black Hole Radiance , \prd 13, 2188-2203, 1976.

\item {}  Wave Function of the Universe , Phys. Rev. D {\bf 28},
2960-2975, 1983. 

\item{} The No-Boundary Measure of the Universe (with T. Hertog),
{\sl Phys. Rev. Lett.},  {\bf 100}, 202301 (2008), arXiv:0711:4630.

\item{} Classical Universes of the No-Boundary Quantum State, (w. T. Hertog), {\sl Phys. Rev. D} {\bf  77}, 123537 (2008), arXiv:0803:1663.

\item{}  The No-Boundary Measure in the Realm of Eternal Inflation (w.  T. Hertog), {\sl Phys. Rev. D}, {\bf 82}, 063510 (2010);  arXiv:1001:0262.

\item{} Local Observation and Eternal Inflation (w.  T. Hertog),  {\sl Phys. Rev. Lett.} {\bf 106}, 141302 (2011);  arXiv:1009.2525.  (An early arXiv version had the title `Eternal Inflation without Metaphysics')

\item{}  Vector Fields in Holographic Cosmology, (w. T.~Hertog),\\ {\sl JHEP11}  (2013) 201,  arXiv:1305.719, 

\item{}  Quantum Probabilities for Inflation from Holography, (w.T.~Hertog)
JCAP, 01(2014) 015,  arXiv:1207.6653.

\item{} Accelerated Expansion from Negative $\Lambda$, (w. T.~Hertog),\\
 arXiv:1205.3807.  
\end{itemize}

\end{document}